\documentclass[aps,twocolumn,pra,superscriptaddress,showpacs,showkeys,floatfix]{revtex4-1}

\usepackage{amsbsy}
\usepackage[pdftex]{graphicx}
\usepackage{caption}
\usepackage{amsmath}
\usepackage{amssymb}
\usepackage{amsfonts}
\usepackage{amssymb}
\usepackage{float}
\usepackage{graphicx}
\usepackage{enumerate}
\usepackage{epstopdf}
\usepackage{braket}
\usepackage{bm}
\usepackage{color,soul}

\begin{document}

\title{Optimal amplification of the dynamical Casimir effect \\
in a parametrically driven system} 

\author{Fabian Hoeb}
\affiliation{Institute for Complex Quantum Systems \& Center for Integrated Quantum Science and Technology (IQST), Ulm University,
  Albert-Einstein-Allee 11, D-89069 Ulm, Germany}
\author{Fabrizio Angaroni}
\affiliation{Center for Nonlinear and Complex Systems,
Dipartimento di Scienza e Alta Tecnologia,
Universit\`a degli Studi dell'Insubria, via Valleggio 11, 22100 Como, Italy}
\affiliation{Istituto Nazionale di Fisica Nucleare, Sezione di Milano,
via Celoria 16, 20133 Milano, Italy}
\author{Jonathan Zoller}
\affiliation{Institute for Complex Quantum Systems \& Center for Integrated Quantum Science and Technology (IQST), Ulm University,
  Albert-Einstein-Allee 11, D-89069 Ulm, Germany}

\author{Tommaso Calarco}
\affiliation{Institute for Complex Quantum Systems \& Center for Integrated Quantum Science and Technology (IQST), Ulm University,
  Albert-Einstein-Allee 11, D-89069 Ulm, Germany}
\author{Giuliano Strini}
\affiliation{Department of Physics, University of Milan,
via Celoria 16, 20133 Milano, Italy}

\author{Simone Montangero}
\affiliation{Theoretische Physik, Universit\"at des Saarlandes, D-66123 Saarbr\"ucken, Germany}
\affiliation{Institute for Complex Quantum Systems \& Center for Integrated Quantum Science and Technology (IQST), Ulm University,
  Albert-Einstein-Allee 11, D-89069 Ulm, Germany}

\author{Giuliano Benenti}
\affiliation{Center for Nonlinear and Complex Systems,
Dipartimento di Scienza e Alta Tecnologia,
Universit\`a degli Studi dell'Insubria, via Valleggio 11, 22100 Como, Italy}
\affiliation{Istituto Nazionale di Fisica Nucleare, Sezione di Milano,
via Celoria 16, 20133 Milano, Italy}
\affiliation{NEST, Istituto Nanoscienze-CNR, 56126 Pisa, Italy}

\begin{abstract}
We introduce different strategies to enhance photon generation in a cavity within the Rabi model in the 
ultrastrong 
coupling regime. 
We show that a bang-bang strategy allows to enhance the effect of up to one order of magnitude with respect to simply driving 
the system in resonance for a fixed time. Moreover, up to about another order of magnitude can be gained exploiting 
quantum optimal control strategies. 
Finally, we show that such optimized protocols are robust with respect to 
systematic errors and noise, paving the way to future experimental 
implementations of such strategies.
\end{abstract}


\maketitle

\section{Introduction}

High-speed manipulation of quantum systems is key
in quantum information processing: Quantum gates should operate on 
a time scale much smaller than the decoherence time, to allow 
efficient error correction and fault-tolerant architectures~\cite{ladd_quant_comp,glaser_cat,caneva_qsl}.
Similarly, the transmission rate is a fundamental characteristic to 
assess the efficiency and the feasibility of real-world application of  
quantum cryptography~\cite{Gisin,Scarani,Lo}. 
Circuit quantum electrodynamics~\cite{blais,wallraff} might play a 
prominent role to speed up quantum protocols, 
since it allows to address the ultrastrong coupling 
regime~\cite{bourassa,gross,mooij,lupascu,semba} 
of light-matter interaction, where the coupling strength 
$\lambda$ becomes comparable to, or even exceeds 
the resonator frequency $\omega$.
This regime has also interesting properties on its own, such as 
the emergence of a strongly correlated light-matter 
ground state~\cite{lupascu,semba}. 

A related problem is the detection of the dynamical Casimir effect
(DCE), namely the generation
of photons from the vacuum due to time-dependent boundary 
conditions or, more generally, as a consequence of the nonadiabatic change of some parameters 
of a system~\cite{moore,dodonov,noriRMP,PA-DCE}.
Indeed, a rapid variation of the matter-field coupling is needed
to implement ultrafast quantum gates, and therefore the DCE appears as a
fundamental limit to the implementation of high-speed quantum
gates~\cite{casimirqip} and more generally to the development of 
ultrafast quantum technologies. 
First experimental demonstrations of the DCE have been 
reported in superconducting circuit quantum electrodynamics~\cite{norinature,lahteenmaki}.
However, it is of great interest to have the ability to either enhance 
or counteract \cite{casimirqip} this 
effect: On the one hand, it improves our capability of investigating fundamental 
effects in nature, and, on the other hand, it enables us to push the limits of
quantum information processing. 

Here, we present three different strategies to amplify and thus improve the 
visibility of the DCE in a parametrically driven system via precisely tailored timing 
of the matter-field interaction. 
In detail, we consider on-off resonance sweeping of a parametrically driven qubit coupled 
to a single mode of the electromagnetic field.
We model the qubit-field interaction by the Rabi 
Hamiltonian~\cite{micromaser}, with a time-dependent modulation of
the qubit frequency.
We use and compare two strategies to optimize the visibility of the DCE:
a multi step heuristic method 
employing bang-bang switches of the qubit frequency from the off-resonance regime
to the resonance regime and back out of resonance, and optimal control theory
which has been proven to be able to successfully control circuit quantum electrodynamics processes~\cite{brif10,glaser_cat,monta07,spoerl07,dong14,goerz16,koch16,allen17}.
In particular, we employ the dressed chopped random basis algorithm (dCRAB) which has been already applied successfully 
to various theoretical and experimental atomic and condensed matter control problems to meet various control goals, including state-transfer, gate synthesis, observable control, and fast quantum phase transition 
crossing~\cite{vanFrank,Brouzos_qsl,caruso12,rosi13,scheuer14,vanfrank14,Caneva_speed}.
For the problem studied here, the control function is the time-dependent modulation of the qubit 
frequency and the figure of merit is the expectation value of 
the number of photons that are generated in the cavity by parametric 
amplification of the DCE: This amounts to finding an optimal setup for the 
detection of the DCE. The results obtained from the dCRAB algorithm are compared 
to the bang-bang strategy and it is shown that 
dCRAB could identify pulses yielding up to about one order of magnitude more 
photons than the bang-bang strategy and up to 
two orders of magnitude more photons than an unoptimized protocol, 
which is a single sweep of the qubit to resonance.

The manuscript is organized as follows: First, we introduce the dynamical system model in Sec.~II. In Sec.~III we 
describe the three control strategies employed to enhance the DCE, discuss the results and give an outlook towards their 
experimental implementation also investigating their robustness against 
some possible experimental imperfections. We conclude our work in Sec.~IV.

\section{The model}

Hereafter, we describe the interaction between a single qubit and a single mode
of the quantized field by means of the Rabi Hamiltonian~\cite{micromaser},
with a time-dependent modulation:
\begin{equation}
\begin{array}{c}
{\displaystyle
H(t)=H_0(t)+H_I,
}
\\
\\
{\displaystyle
H_0(t)=-\frac{1}{2}\,[\omega_{q0}+\dot{\Phi}(t)] \sigma_z +
\omega\left(a^\dagger a +\frac{1}{2}\right),
}
\\
\\
{\displaystyle
H_I=\lambda \,\sigma_+\,(a^\dagger+a)
+\lambda^\star \sigma_-\,(a^\dagger+a),
}
\end{array}
\label{eq:noRWAparam}
\end{equation}
where the reduced Planck's constant is set to $\hbar=1$, 
$\omega_{q0}$ being the reference frequency for the qubit,
$\sigma_i$ ($i=x,y,z$) are the Pauli matrices,
written in the $\{|g\rangle,|e\rangle\}$ basis;
$\sigma_\pm = \frac{1}{2}\,(\sigma_x\mp i \sigma_y)$
are the raising and lowering operators for the qubit
(so that $\sigma_+=|e\rangle\langle g|$ and
$\sigma_-=|g\rangle\langle e|$):
$\sigma_+ |g\rangle = |e\rangle$,
$\sigma_+ |e\rangle = 0$,
$\sigma_- |g\rangle = 0$,
$\sigma_- |e\rangle = |g\rangle$.
The operators $a^\dagger$ and $a$ for the field create
and annihilate a photon:
$a^\dagger |n\rangle=\sqrt{n+1}|n+1\rangle$,
$a |n\rangle=\sqrt{n}|n-1\rangle$,
$|n\rangle$ being the Fock state with $n$ photons.
For the sake of simplicity, we consider a real 
coupling strength, $\lambda\in\mathbb{R}$.
The real function $\dot{\Phi}(t)$ is the control field which allows to manipulate 
the system: the qubit frequency is modulated via $\omega_q(t)=
\omega_{q0}+\dot{\Phi}(t)$. Notice that we used the notation with the first derivative because, as
it will become apparent in the next paragraph, the relevant quantity 
is the accumulated phase, i.e. ${\Phi}(t)$. 

The Rotating Wave Approximation (RWA) 
(valid for $\lambda\to0$)
is obtained neglecting the term
$\sigma_+ a^\dagger$, which simultaneously
excites the qubit and creates a photon,
and $\sigma_- a$, which de-excites the
qubit and annihilates a photon. In this limit, the Hamiltonian
(\ref{eq:noRWAparam}) reduces to the Jaynes-Cummings
Hamiltonian \cite{micromaser} with a time-dependent
modulation.
In the RWA the swapping time needed to transfer
an excitation from the qubit to the field or vice versa
($|e\rangle |0\rangle\leftrightarrow |g\rangle |1\rangle$)
is $\tau_s=\pi/2\lambda$, {and no DCE is possible since the 
total number of excitations in the system is conserved}. 

In the interaction picture, we first consider 
the unitary operator 
\begin{equation}
\begin{array}{c}
{\displaystyle
U(t)=\mathcal{T}\exp\left[-i\int_{0}^t d\tau \,H_0(\tau)\right]
}
\\
{\displaystyle
=\mathcal{T}\exp\left\{\frac{i}{2}\,[\omega_{q0}t+\Phi(t)]\,\sigma_z
-i\omega t\left(a^\dagger a+\frac{1}{2}\right)\right\},}
\end{array}
\end{equation}
where $\mathcal{T}$ is the time-ordering operator and 
$\Phi(t)=\int_0^td\tau\,\dot{\Phi}(\tau)$
the accumulated phase. 
The time-dependent Hamiltonian in the interaction picture then reads
\begin{equation}
\begin{array}{c}
{\displaystyle
\tilde{H_I}(t)=U^\dagger(t) H_I U(t)
}
\\
{\displaystyle
=\lambda\left(
a\sigma_- \exp\{-i[(2\omega-\Delta_0)\,t+\Phi(t)]\}
\right.
}
\\
{\displaystyle
\left.
+a\sigma_+ \exp\{i[-\Delta_0\, t+\Phi(t)]\}
\right.
}
\\
{\displaystyle
\left.
+a^\dagger\sigma_- \exp\{-i\,[-\Delta_0\, t+\Phi(t)]\}
\right.
}
\\
{\displaystyle
\left.
+a^\dagger\sigma_+ \exp\{i\,[(2\omega-\Delta_0)\,t+\Phi(t)]\}
\right),
}
\end{array}
\end{equation}
where we have defined the reference detuning 
$\Delta_0=\omega-\omega_{q0}$.
The standard Rabi model is recovered for the time-independent 
Hamiltonian, $\Phi(t)=0$, and the Jaynes-Cummings model if we further
neglect the counter-rotating terms at frequency $2\omega$. 
From now on we will omit tildes and always refer to the interaction 
picture. 

If we expand, in the interaction picture, 
the qubit-field state at time $t$ as
$|{\Psi}(t)\rangle=\sum_{l=g,e}\sum_{n=0}^\infty
{C}_{l,n}(t) |l,n\rangle$, we obtain the equations 
that govern the evolution of the coefficients ${C}_{l,n}(t)$:
\begin{equation}
  \left\{
\begin{array}{c}
{\displaystyle
    i  \,\dot{C}_{g,n}(t)  =
    \Omega_n e^{-i\,[-\Delta_0\, t + \Phi(t)]}\,{C}_{e,n-1}(t) 
}
\\
{\displaystyle
 + \Omega_{n+1} e^{-i\, [(2\omega-\Delta_0)\, t+\Phi(t)]} \,C_{e,n+1}(t),
}
  \\
\\
{\displaystyle
    i  \,\dot{C}_{e,n}(t)  =
    \Omega_{n+1} e^{i\,[-\Delta_0\, t + \Phi(t)]}\,{C}_{g,n+1}(t) 
}
\\
{\displaystyle
      + \Omega_{n} e^{i \,[(2\omega-\Delta_0)\, t+\Phi(t)]} \,C_{g,n-1}(t).
}
  \end{array}\right.
  \label{eq:coefficients}
\end{equation}
Here $\Omega_n=\lambda\sqrt{n}\,$ are the Rabi frequencies,
with $n=0,1,2,...$ (the terms $C_{l,m}$ and $\dot{C}_{l,m}$ must be set to zero
when $m<0$).
In numerical simulations we will set the reference detuning $\Delta_0=0$ to ensure that the off resonance condition $|\omega - \omega_q| \ll \lambda$ holds in the ultrastrong coupling regime~\cite{bourassa}. 

\section{Results}

We consider the time evolution of the qubit-oscillator system 
for an overall time interval $T$.
An initial and a final time interval $\tau$,  
corresponding to off-resonance evolutions with detuning
$|\omega - \omega_q(t)| \gg \lambda$, are excluded from manipulation to determine the figure of merit. 
Consequently, on-off resonance sweeps (governed by a time-dependent pulse
$\omega_q(t)$) are possible during 
the intermediate time span $\tau_p= T- 2\tau$. 
Initially both the field and the qubit are prepared in their ground 
state $|\Psi(t=0)\rangle=|g,0\rangle$,
 so that within the RWA there is no generation of photons at all times ($\langle n\rangle=0 \; \forall t$).
On the other hand, when the terms beyond the RWA are taken into account,
only a very weak photon generation is possible in the off-resonance
regime, while a significant photon emergence is possible if the 
resonance condition is approached nonadiabatically.
To quantify the strength of this manifestation of the DCE effect, we 
consider the figure of merit
\begin{equation}
f=\bar n_f - \bar n_i, \label{eq:fom}
\end{equation}
where $\bar n_{f/i} = \int \langle n(t) \rangle dt$ 
is the time-average of the mean photon number $\langle n \rangle$
over the initial (i) and final (f) off-resonance time intervals of duration $\tau$ each. 
Hereafter, we set the off-resonance condition to $\omega_q=4\omega$; the exact numerical value has, however, a very minor influence only.

\subsection{On-off protocol}

We begin by studying the efficiency of a straightforward protocol, namely
a sudden switch to resonance $\omega_q=\omega$ at time $\tau$, 
followed by another instantaneous quench to the off-resonance condition at time 
$T-\tau$. 
That is, the system is kept on resonance for a total time $\tau_p$. 
 
\begin{figure}[t]
\vskip 1cm
\includegraphics[angle=0.0, width=8cm]{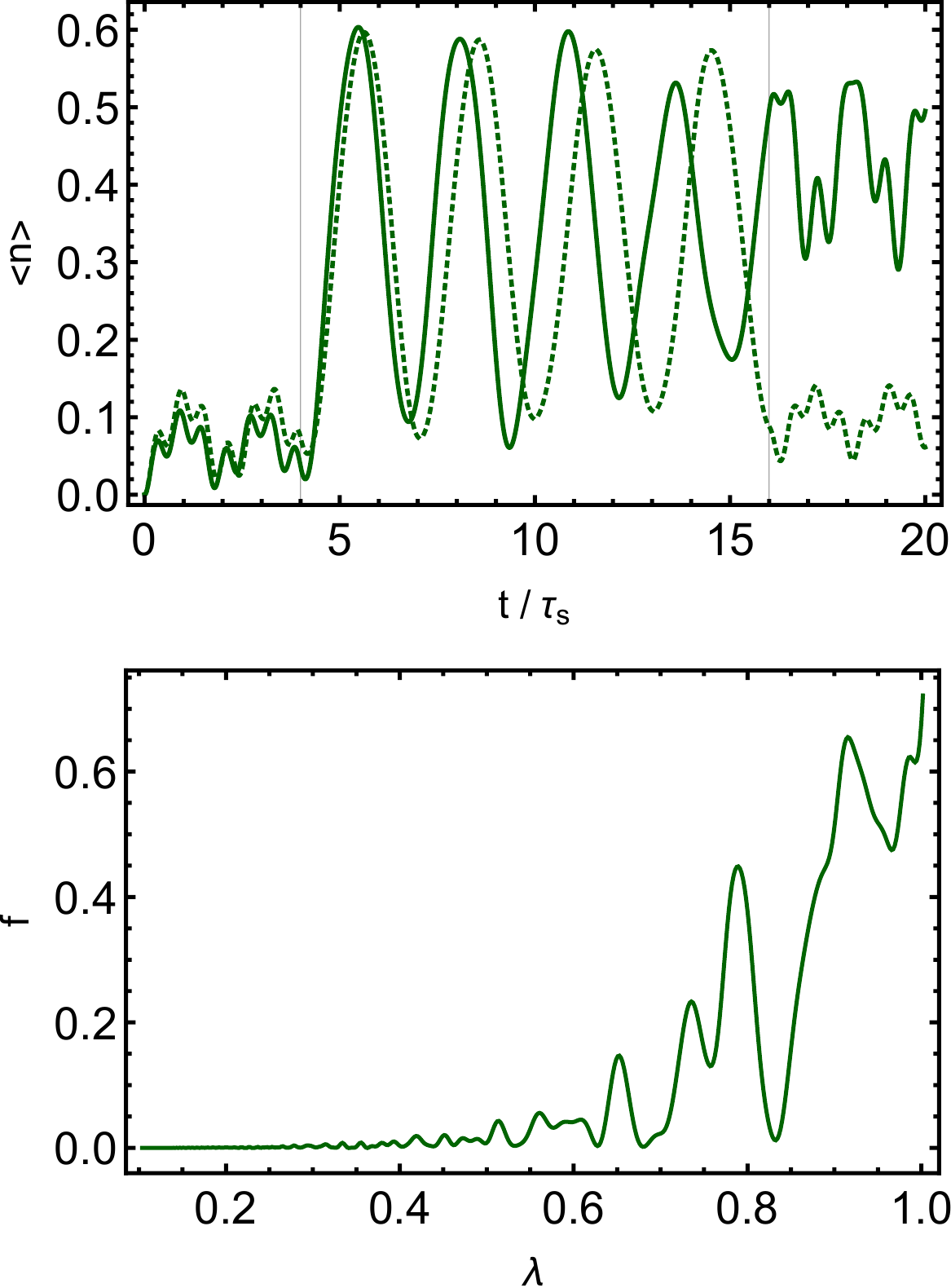}
\caption{
Results for the on-off protocol.
Top: Dynamical evolution of the mean number of  
photons $\langle n \rangle$
at interaction strength 
$\lambda=0.80$ (solid green line) and $\lambda=0.83$ 
(dashed green line).
Bottom: Figure of merit $f$ 
as a function of the coupling strength $\lambda$.
For both figures, the initial and final off-resonance periods, as indicated by the 
vertical black lines, are set to $\tau=4\tau_s$.}  
\label{fig:dynamics}
\end{figure}

Typical examples of time evolutions of the instantaneous average number 
of generated photons $\langle n \rangle$ are shown in Fig.~\ref{fig:dynamics} (top),
in the ultra-strong coupling regime, for $\lambda=0.80$ (solid line) and 
$\lambda=0.83$ (dashed line): it is clearly visible that the number of photons 
remains quite small in the initial off-resonance regime ($t< \tau$), while 
$\langle n \rangle$ grows rapidly after switching to the resonance condition
(at time $t=\tau=4\tau_s$). 
It can be clearly seen that at resonance $\langle n \rangle$ 
does not grow indefinitely, but oscillates due to coherent generation 
(dynamical Casimir effect) and destruction of photons (anti-dynamical
Casimir effect \cite{antiDCE}). Finally, after the switch 
to the off-resonance regime (at time $t=T - \tau=16\tau_s$), the 
average photon number keeps on oscillating
around its value at $t=T-\tau$ with smaller amplitude oscillations compared to the ones on resonance. 
In Fig.~\ref{fig:dynamics} (bottom) the results of this protocol are summarized, reporting the value of 
the figure of merit $f$ given by Eq.~\eqref{eq:fom} versus $\lambda$.
Note that this protocols reveals a strong sensitivity with respect to 
variations of the system parameters, especially in the regime $\lambda \gtrsim 0.6$ where a non negligible number of 
photons in the cavity are generated. 
Indeed, a slight change of the coupling strength $\lambda$ induces a strong
variation of the mean number of photons: 
For instance, in the cases reported in Fig.~\ref{fig:dynamics} (top)  
we obtain $f \approx 0.37$
for $\lambda=0.80$ whereas $f \approx 0.016$ for $\lambda=0.83$.
Such a strong sensitivity suggests that accurate control of the 
system parameters is needed for reliable implementation 
of quantum protocols in the ultrastrong coupling regime.

\subsection{Optimized strategies}

In order to amplify and improve the visibility of the DCE, 
more sophisticated pulses with multiple on-off resonance sweeps are required. 
Hereafter we will implement and compare two methods, a bang-bang strategy~\cite{bangbang} and 
optimal control using dCRAB~\cite{CRAB,dCRAB}.

For the bang-bang strategy, the qubit is placed in-out resonance with 
instantaneous sweeps. The main idea of this strategy is pretty straightforward: the duration of the on-resonance 
intervals ($\omega_q=\omega$) is determined by the time 
needed to reach the first maximum in the number of photons, while the detuned intervals ($\omega_q=4\omega$) last for a fixed time $ \tau_O \ll \tau_p$ since gain in $f$ is mainly seen on resoncance.  
This strategy already provides an improvement with respect to the on-off protocol (data not shown), however we are 
going one step further and employ a more sophisticated strategy, which makes use of multiple iterations steps, to obtain best possible on-resonance time interval 
lengths
(for details see Appendix A).  

The second approach to target the objective $f$ is by means of the dCRAB optimal control technique~\cite{dCRAB}.  
The two main ingredients of dCRAB are an expansion of the control functions over a randomized truncated basis, 
and iterative re-initialization of local searches allowing the algorithm to escape from false traps~\cite{CRAB,dCRAB}:
The basis functions, which define the subspace subject to search, are updated at each re-initialization step and the new 
emerging 
search directions are explored by means of gradient-free minimization algorithms.\\

\begin{figure*}[t]
\vskip 1cm
\includegraphics[angle=0.0, width=17cm]{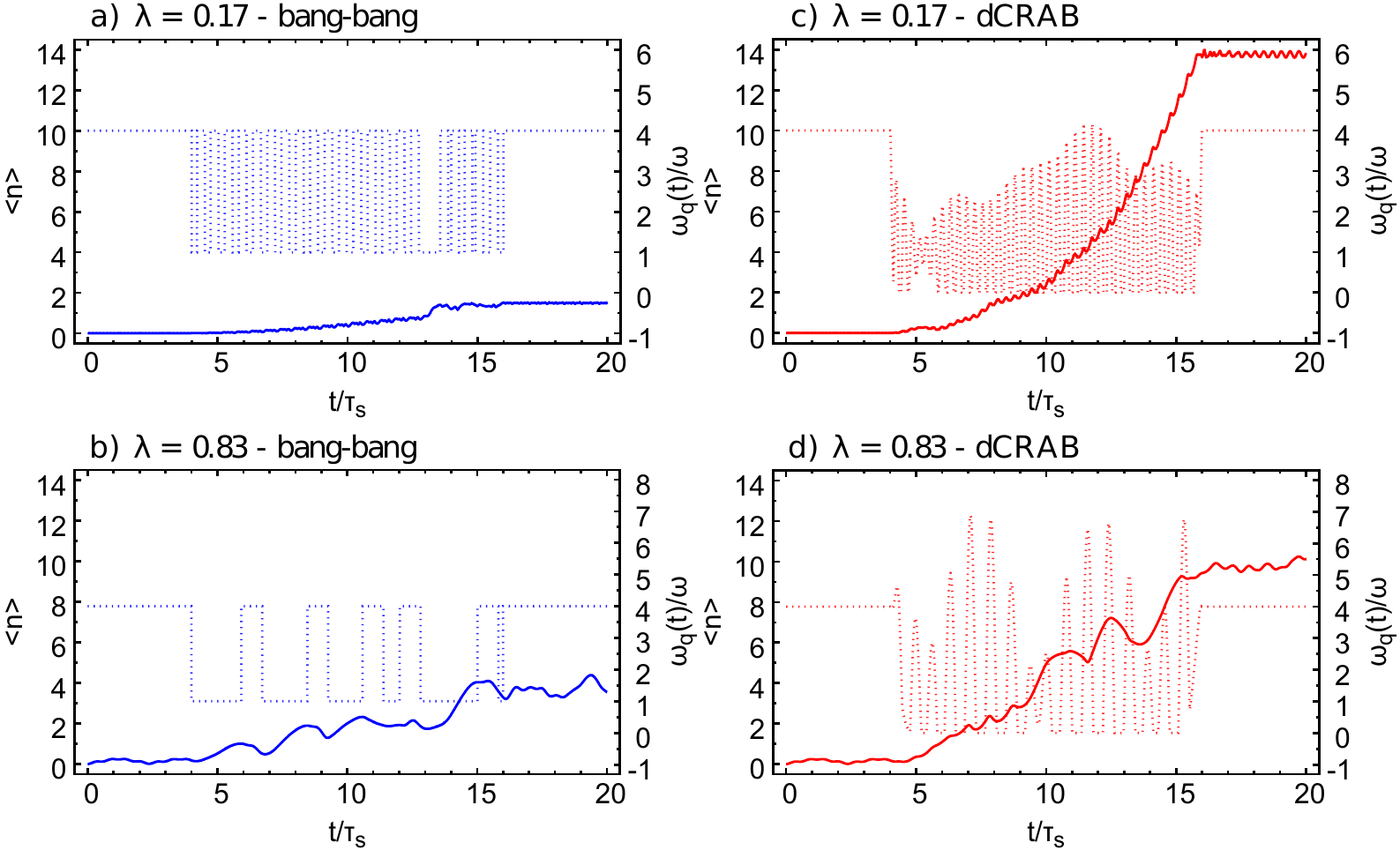}
\caption{
Expected photon number $\langle n \rangle$ versus time for bang-bang strategy (left, blue solid lines) and dCRAB 
(right, red solid lines) for coupling strengths $\lambda = 0.17$ (a,c) and $\lambda = 0.83$ (b,d). On the secondary 
y-axis, the 
control pulses (dashed lines, colors respectively) are displayed. Note that there is an initial and final time interval 
equally out of 
resonance for both strategies,
used to compute the figure of merit $f$ (see Eq.~\eqref{eq:fom}).}  
\label{fig:res83}
\end{figure*}

Two representative results for coupling strengths $\lambda = 0.17$ ($\lambda = 0.83$) are presented in 
Fig.~\ref{fig:res83}a,c (Fig.~\ref{fig:res83}b,d) 
comparing the dCRAB solution to the outcome of the bang-bang strategy: the 
expectation values of the photon number are reported along with the generating control pulses 
$\omega_q(t) / \omega$. From these exemplary cases, the working principle of the 
bang-bang strategy is clearly revealed: Once the photon number peaks to a new maximum, the qubit frequency is 
driven out of resonance to let the system settle for some pre-defined time until it is switched back on resonance (see, e.g., 
Fig.~\ref{fig:res83}b).
On the contrary, for the dCRAB solution the interplay between $\langle n \rangle$ and the control input 
is no longer apparent. 
However, it can be seen that, after some time, the number of photons generated via the dCRAB pulse always exceeds 
the bang-bang photons number (this is true for all cases tested). Notice that the 
non-negativity condition ($\omega_q(t) \ge 0$) for the pulses is imposed at all times, 
preventing a swap in the computational basis of the qubit. 

\begin{figure}[htp]
\vskip 1cm
\includegraphics[angle=0.0, width=8.5cm]{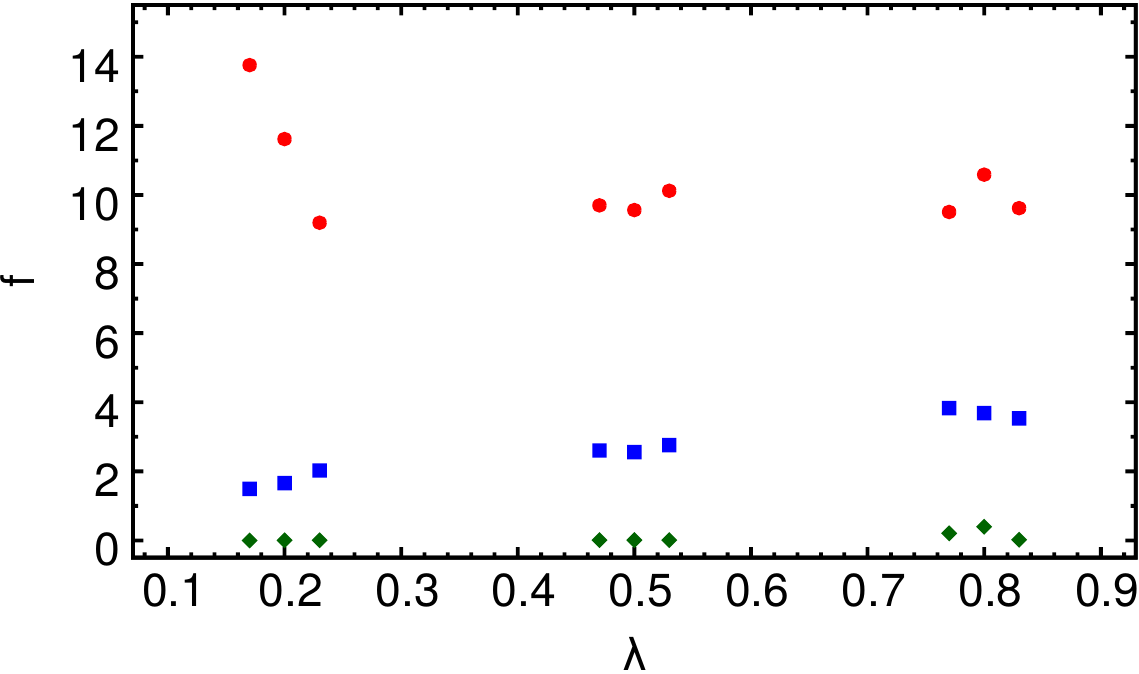}
\caption{
Created photon number $f$ versus coupling strength $\lambda$ after application of three different control strategies: 
dCRAB optimization (red circles), bang-bang strategy (blue squares) and unoptimized on/ off resonance sweeps (green 
diamonds). Note that the dCRAB optimized solutions improve significantly compared to the other two strategies and hence 
pave the way towards optimal amplification of the DCE. 
}  
\label{fig:fom3}
\end{figure}

The results for different coupling strengths $\lambda$ are summarized in Fig.~\ref{fig:fom3}, where the protocol's 
performances are
clearly visible. Overall, in the ultrastrong coupling regime the improvement from the dCRAB optimizations over the bang-bang strategy is of up 
to one order of magnitude; compared to the unoptimized single on-off resonance protocol, the optimal solutions 
yield up to two orders of magnitudes more photons.
We also tested the regime with coupling $\lambda\ll 1$:
For $\lambda=0.03$, dCRAB 
leads to a figure of merit $f \approx 12$, whereas 
the other two strategies perform very poorly ($f$ well below 0.2). 
Quite interestingly, the dCRAB strategy shows that amplification of the DCE with 
a significant photon production is also possible for $\lambda\ll 1$, 
at the price of a longer duration of the protocol, 
$T=20\tau_s\propto 1/\lambda$.  


\begin{figure}[t]
\vskip 1cm
\includegraphics[angle=0.0, width=8cm]{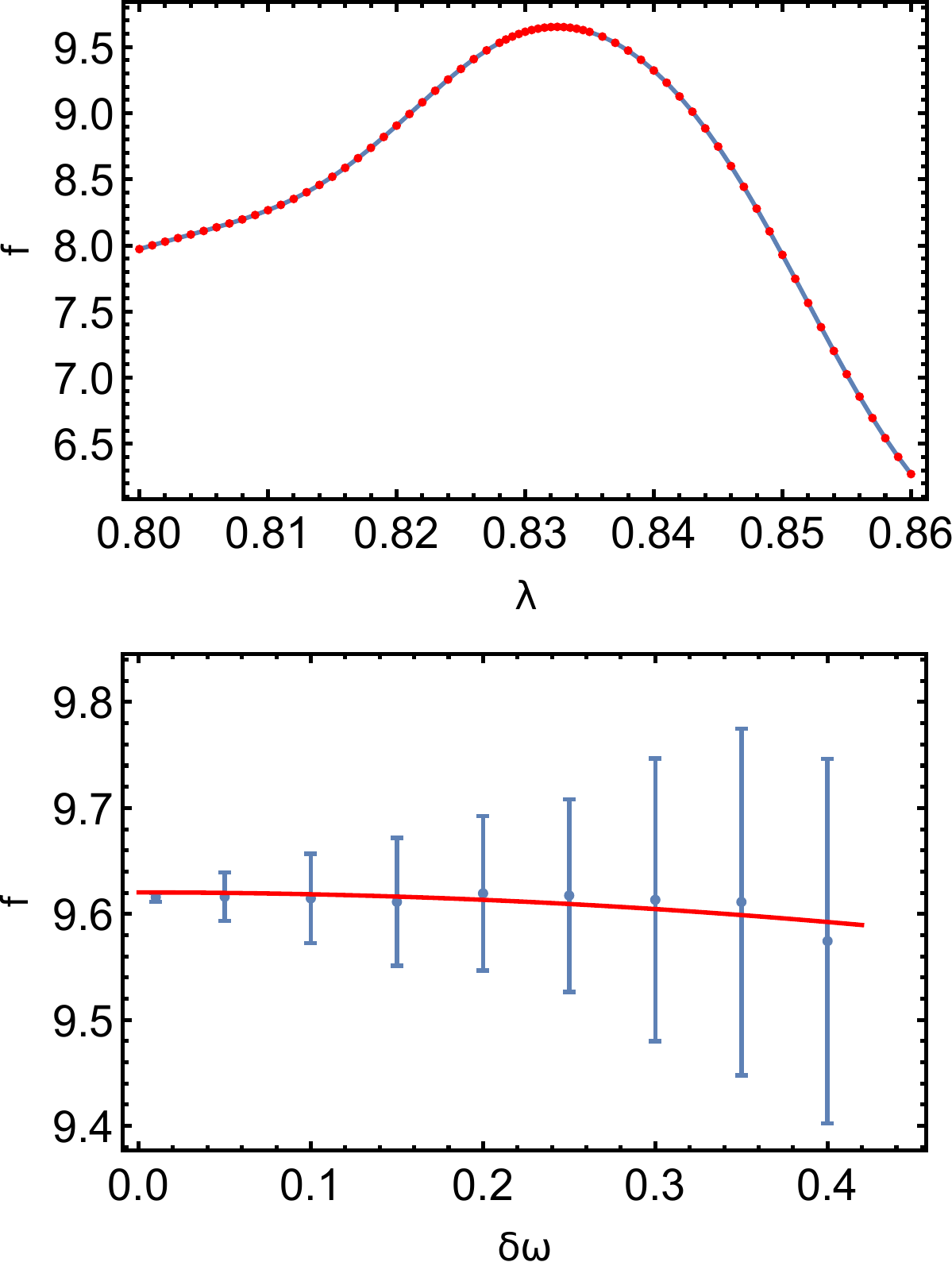}
\caption{
Figure of merit for the dCRAB pulse computed for the unperturbed $\lambda=0.83$ case,
when the system is exposed either to a systematic (top) or 
statistical (bottom) errors. In the former case, the actual 
value of $\lambda$ set for the propagation ranges in (0.80, 0.86). In the latter, the qubit frequency 
$\omega_q(t)$
is affected by random noise uniformly distributed in the interval
$[-\delta\omega,\delta\omega]$, with correlation time 
$\tau_c\approx 0.03 \; T$. That is, $\xi(t)$ is  
reset every time interval $\tau_c$. Error bars denote statistical
errors for 100 realizations of noise per point, while the red straight line shows the outcome from  
a quadratic fit.}
\label{fig:errors}
\end{figure}

Finally, we point out that the optimal pulses identified here are 
expected to work equally well in a (reasonable) noisy environment, due to their intrinsic robustness against small variations, as it has been already theoretically and experimentally showed in many different scenarios~\cite{monta07,lloyd14,Frank_nv_cntrl,Laflamme_nmr,biercuk_prediction_decoherence,kosut13}. 
Moreover, if closed loop optimal control is employed, the optimization incorporates unknown and unpredictable drifts 
into the pulse design as well as makes the pulses robust against statistical disturbances (noise on the pulses and 
the figure of merit)~\cite{rosi13,Frank_nv_cntrl,Laflamme_nmr}.
Indeed, we could confirm the robustness of the optimal strategies by numerical simulations of the system 
evolution steered by the optimized dCRAB pulse (from Fig. 3d) and additionally affected by either systematic or 
statistical errors. 
In the former case, we assume the presence of a systematic error in the coupling strength compared to one specific 
coupling strength that was used in the optimization. For the analysis shown in Fig.~\ref{fig:errors} (top), we choose 
$\lambda=0.83$ as the reference strength for the optimization and show the outcome $f$ for different values of $\lambda \in 
(0.80,0.86)$. We can conclude that for the tested range, the figure of merit $f$ still remains reasonably good (more than 
2/3 of the optimized $f$), and anyway much bigger than what could be obtained with the other strategies. 
Moreover, notice that the region of this particular coupling strength is highly sensitive in terms of 
photon generation as we could see for the on-off protocol in Fig.~\ref{fig:dynamics}: For instance, going from 
$\lambda=0.80$ to 
$\lambda=0.83$, means a drop in $f$ by a factor of more than 20 for the unoptimized protocol. 

Finally, we analyze the scenario where the optimal pulse $\omega_q(t)$ is affected by random noise $\xi(t)$
uniformly distributed in the interval 
$[-\delta \omega,\delta\omega]$. 
In Fig.~\ref{fig:errors} (bottom),
we can see that the figure of merit $f$ (averaged over 
100 noise realizations) is very stable in the range 
$\delta\omega/\omega=[0,0.4]$ and was fitted to scale as $f(\delta\omega) = (9.62 \pm 0.00494) - (0.174 \pm 0.0633) 
\cdot \delta\omega^2$. The correlation time of the noise was set to be $\tau_c = 0.03 \,T$, however we checked over about 
three orders of magnitude that it does have an almost negligible effect only (data not shown). 

\section{Conclusions}

We have applied different optimization strategies
to a driven Rabi model Hamiltonian
in the ultrastrong coupling regime. In particular, we have optimized 
the number of photons generated by the dynamical Casimir effect, 
in order to enhance its visibility in view of possible future experimental verifications. 
The huge amplification of the DCE given by the dCRAB optimization 
is quantified  by an enhancement 
of the generated photons of up to about an order of magnitude 
with respect to a reference bang-bang strategy and of up to about two 
orders of magnitude with respect to the unoptimized
on-off strategy. 

The optimization performed in this paper is a valid proof
of concept. Moreover, the parameters used in our simulations 
are within reach for present technology, where coupling strengths even
exceeding the resonator frequency have been recently 
reported~\cite{lupascu,semba}. Additionally, in circuit quantum electrodynamics experiments
the populations of Fock and coherent states were measured by means 
of the Fourier transform of the time-dependent polarization signal 
of a probe qubit interacting with the field~\cite{martinis};
extensions of this method~\cite{tomography} 
would allow not only the detection of the number of 
photons but also the reconstruction of the exotic 
field states generated by the DCE~\cite{exotic}.
More realistic models would need a detailed treatment of decoherence
sources and experimental imperfections, e.g. in driving the matter-field
coupling or measuring the final state of the system. Hence, given that 
our results have been obtained for an idealized scenario, they are 
an estimate of the upper bound of the strength of the DCE in the driven Rabi model. 
However, as we have shown that optimal solutions are robust with respect to noise sources 
and systematic errors, we are optimistic about the possible experimental verification 
of the presented results in the future. 

Finally, the same approach can be used to prepare and investigate different states and 
phenomena in the ultrastrong coupling regime, as for example, targeting  
a given Fock state of the field or generating squeezed field states. 

We acknowledge support from the EU via the RYSQ project and from the DFG via the SFB/TRR21 and from BMBF via the project Q.Com. 
S.M. gratefully acknowledges the support of the DFG via a Heisenberg fellowship.

\appendix
\section{Optimized Bang-bang strategy}

\begin{figure*}[htb]
\begin{center}
\includegraphics[width = 17cm]{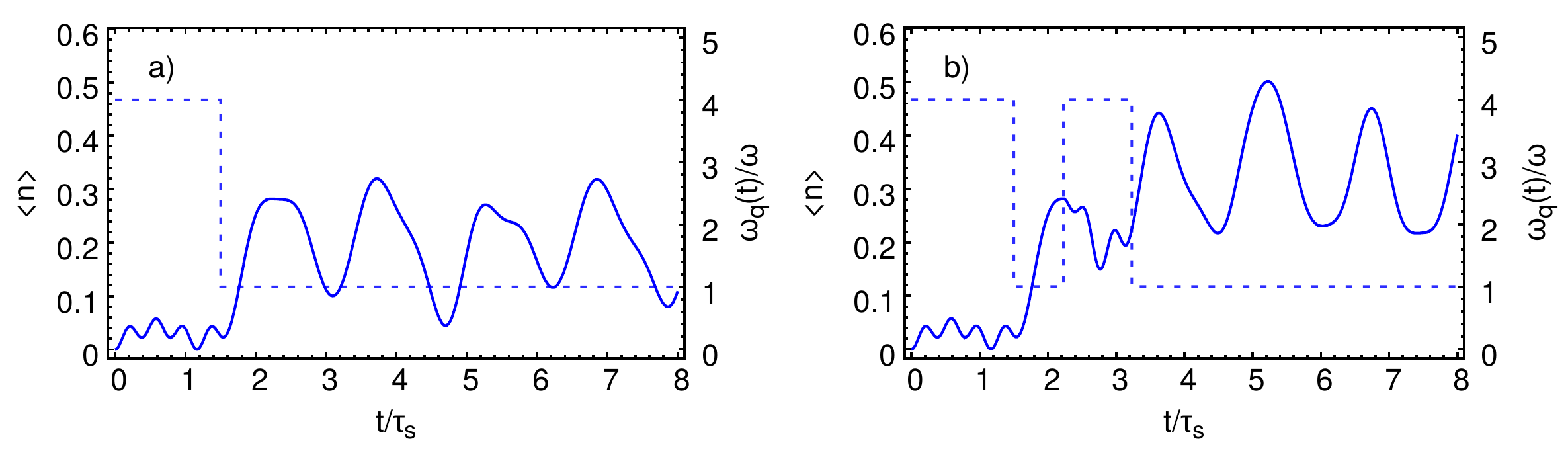}
\caption{Working principle of the bang-bang strategy shown for an initial pulse (a) and the first iteration 
of the algorithm (b), for 
$\lambda = 0.5$ and $\tau_O/\tau_s =1.5$.
During the time on resonance of the pulse (dashed line in a)) 
$\Braket{n}(t)$ oscillates around an average value (solid line in a)). Upon switching back out of 
resonance at the maximum of the first oscillation (the corresponding pulse is the dashed line in b)), the values of 
$\Braket{n}$ remains almost frozen around this maximum (solid line in b)). Reverting back to the resonance regime  
leads to the oscillatory behavior of $\Braket{n}$. 
The initial fast increase  after each switch on resonance results in the overall population increase of the cavity.
}
\label{fig: ref_strat_principle}
\end{center}
\end{figure*}

We build optimized pulses employing bang-bang switches from an off resonance regime, 
where $\omega_q = 4\omega$, to the resonance regime $\omega_q = \omega$ and back out of resonance after some fixed time $\tau_O$. 
The pulses have a total duration of $T = 20 \cdot \tau_s$, however we reserve an initial and final $\tau$ where the pulse is kept constant to compute 
$\bar n_{f/i}$ to determine the figure of merit $f$ in Eq. (5).

To optimize the pulses the system is firstly evolved according to an initial single bang-bang pulse 
$\omega_{q,ref}^{(0)}$ (dashed line in Fig. \ref{fig: ref_strat_principle} a)), which is iteratively improved according 
to the following procedure:
\begin{description}
\item[1)] Evaluate the time evolution for the pulse $\omega_{q,ref}^{(i)}$ and locate the first local maximum of photon number $\langle n_\textrm{M}(t_M^{(i)}) \rangle$ after the last switch to resonance in the current pulse.  
\item[2)] Generate a new pulse $\omega_{q,ref}^{(i+1)}$ based on $\omega_{q,ref}^{(i)}$ by switching out of 
resonance at $t_M^{(i)}$ (point in time of first occurring maximum in considered time interval). Remain out of 
resonance 
for some initially specified time $\tau_O$. Finally, at $t_M^{(i)} + \tau_O$, the pulse 
reverts back to resonance for the remaining $t \in (t_M^{(i)} + \tau_O,T-\tau)$. 
\item[3)] At $t=T-\tau$ switch back off-resonance to allow for the computation of the figure of merit $f$. 

\item[4)] Iterate steps 1) - 3) and stop when $t_M^{(i)} > T-\tau$. 

\end{description} 
In Fig.~\ref{fig: ref_strat_principle} we show a typical result of the first iteration of this strategy: It can be 
clearly seen that after going on resonance the photon number rapidly increase and stabilities to an higher value. The 
introduction of an non negligible off-resonance time period $\tau_O$ leads to generally better results by letting the 
system equilibrate to the new conditions.

\newpage



\begin{thebibliography}{0}
\bibitem{PA-DCE}
The phenomena where the photons are created by parametric amplification 
of vacuum fluctuations without moving boundaries is referred to as 
parametic DCE \cite{dodonov}. This is the case for the driven Rabi Hamiltonian 
discussed in this paper.
\end{thebibliography}
\end{document}